\documentclass[twocolumn,preprintnumbers,showpacs]{revtex4-1}
\usepackage{amsmath}
\usepackage{amssymb}
\usepackage{graphicx}
\usepackage{color}
\usepackage{float}
\usepackage{sistyle}
\usepackage{subfigure}  
\usepackage{verbatim}   
\usepackage{graphicx}
\usepackage{dcolumn}
\usepackage{bm}
\usepackage[citecolor=blue,filecolor=blue,colorlinks=true,urlcolor=blue]{hyperref}
\usepackage{ulem}

\newcommand{\E}{\ensuremath{E_{g}\,}}
\newcommand{\AO}{\ensuremath{A_{1g}\,}}
\newcommand{\BAN}{\ensuremath{B_{1g}\,}}
\newcommand{\BN}{\ensuremath{B_{2g}\,}}

\graphicspath{{images/}{../images/}}

\begin{document}

\title{Observing the Superconducting State of HgBa$_2$Ca$_2$Cu$_3$O$_{8+\delta}$ Cuprate by Electronic Raman Scattering}

\author{B. Loret$^1$, A. Sacuto$^1$, D. Colson$^2$, Y. Gallais$^1$, M. Cazayous$^1$, M.-A. M\'easson$^1$ and A. Forget$^2$}

\affiliation{$^1$ Laboratoire Mat\'eriaux et Ph\'enom$\grave{e}$nes Quantiques (UMR 7162 CNRS), Universit\'e Paris Diderot-Paris 7, Bat.Condorcet, 75205 Paris Cedex 13, France,\\
$^2$ Service de Physique de l'{\'E}tat Condens{\'e}, DSM/IRAMIS/SPEC (UMR 3680 CNRS), CEA Saclay 91191 Gif sur Yvette cedex France.}

\date{\today}


\begin{abstract}
Electronic Raman scattering with in and out of (ab) plane polarizations have been performed on HgBa$_2$Ca$_2$Cu$_3$O$_{8+\delta}$ in a slightly underdoped single crystal with a critical temperature T$_{c}$=122 K. We find that the $d$-wave pairing gap at the antinodes is higher in energy ($14~k_{B}$T$_{c}$) than in other cuprates and that it varies very slowly up to $T_{c}$. This hints at a strong coupling nature of the pairing mechanism. Interestingly, we reveal that  the pairing-gap feature in the Raman response displays a complex peak-dip-hump structure, in a fashion reminiscent of what observed by angle resolved photo-emission spectroscopy in Bi$_2$Sr$_2$CaCu$_2$O$_{8+\delta}$ (Bi-2212). We detect two other distinct superconducting peaks at $\approx5k_{B}$T$_{c}$ and $\approx7k_{B}$T$_{c}$ when probing respectively around the nodes and on the whole Fermi surface. Finally we establish that the pairing gap at the antinodes is detected both for (ab) plane and for c-axis light polarizations. This shows that the quasiparticle dynamics along the c-axis is intimately connected to the antinodal one in the (ab) plane. 

\end{abstract}

                                                     
\maketitle
\section{Introduction}

The discovery of the three-copper-oxide-layer compound HgBa$_2$Ca$_2$Cu$_3$O$_{8+\delta}$ (Hg-1223) by  A. Schilling et al.\cite{Schilling1993} in 1993 was a breakthrough in the quest of high critical temperature superconductors. Indeed, the Hg-1223 cuprate exhibits a critical temperature $T_{c}$ of 135~K at ambient pressure, i.e.  $40$ K higher than $T_{c}$ of YBa$_{2}$Cu$_3$O$_{7-\delta}$, and up to $164~K$ under 30 GPa pressure\cite{Gao1994}. However very few spectroscopic studies have been carried out on this material \cite{Ren1993,Yang1994,Sacuto1996,Zhou1996,Zhou1997,Sacuto1997,Sacuto1998,Wei1998,Sacuto2000a,McGuire2000}  due to the difficulty to grow large enough single crystals with high optical quality surface. 
Working on the highest-$T_{c}$ cuprate superconductor could therefore give us the chance to understand what gives rise to such a high $T_{c}$. Although the earlier Raman measurements on Hg-1223 suggested a multicomponent anisotropic superconducting (SC) gap (with two distinct gap maxima) \cite{Sacuto1997,Sacuto1998,Zhou1997bis}, it turned out that the experimental findings were more appropriately described by a $d$-wave pairing gap \cite{Wei1998,Sacuto2000a,McGuire2000}. The d-wave SC gap amplitude is expected to be maximum near the principal axes of the Brillouin zones ($(\pm\pi,0)$ and $(0,\pm\pi)$), called the anti-nodal regions, while it is expected to vanish near the diagonal of the Brillouin zone $(\pm\pi/2,\pm\pi/2)$, called the nodal regions \cite{Devereaux2007,Sacuto2013}. 

Here we present new electronic Raman scattering studies on Hg-1223, which greatly improve our previous results (fifteen years ago) \cite{Sacuto1996,Sacuto1997,Sacuto1998,Sacuto2000a} by investigating the Raman spectra over a significantly wider energy range and revealing spectral features previously overlooked. We have performed both in and out of plane polarization Raman study on slightly under-doped Hg-1223 (T$_{c}$=122 K). In the (ab) plane we focus on the \BAN, \BN and \AO symmetries, which allow us to probe respectively the anti-nodal region, the nodal region and the whole Brillouin zone in the ($k_{x},k_{y}$) plane. We find that the maximum amplitude of the pairing gap was underestimated in our previous work \cite{Sacuto1996,Sacuto1997,Sacuto1998,Sacuto2000a}, resulting much higher in energy at $1135~cm^{-1}$ ($14~k_{B}$T$_{c}$) instead of $800~cm^{-1}$ ($9~k_{B}$T$_{c}$). 

We reveal a weakly temperature-dependence of the pairing gap $2\Delta$ on a large temperature-range up to $T_{c}$. We also show that the pairing gap is associated with a peak-dip-hump structure in the Raman response, similarly to what observed by angle resolved photo-emission spectroscopy (ARPES) in Bi-2212 \cite{Fedorov1999,Campuzano1999}. 
Moreover we find two other distinct energy scales in \BN symmetry and \AO symmetry which disappear above $T_{c}$. 
The energy ratio between the \BAN and \BN peaks is extremely high $\approx2.6$ in comparison with what we expect for a standard $d$-wave SC gap ($\approx1.3$) \cite{Devereaux1995}.
Using light polarizations along the c-axis we have also probed the \AO and \E symmetries.
Interestingly, we detect in \AO symmetry the same pairing peak at 1135 $cm^{-1}$ as the one found in the (ab) plane with the \BAN symmetry. This shows  that the c-axis electronic properties and the antinodal ones within the (ab) plane are tightly connected. 

\section{Details of the Experimental Procedure}

\begin{figure}[htp!]
\begin{center}
\includegraphics[width=6cm,height=4cm]{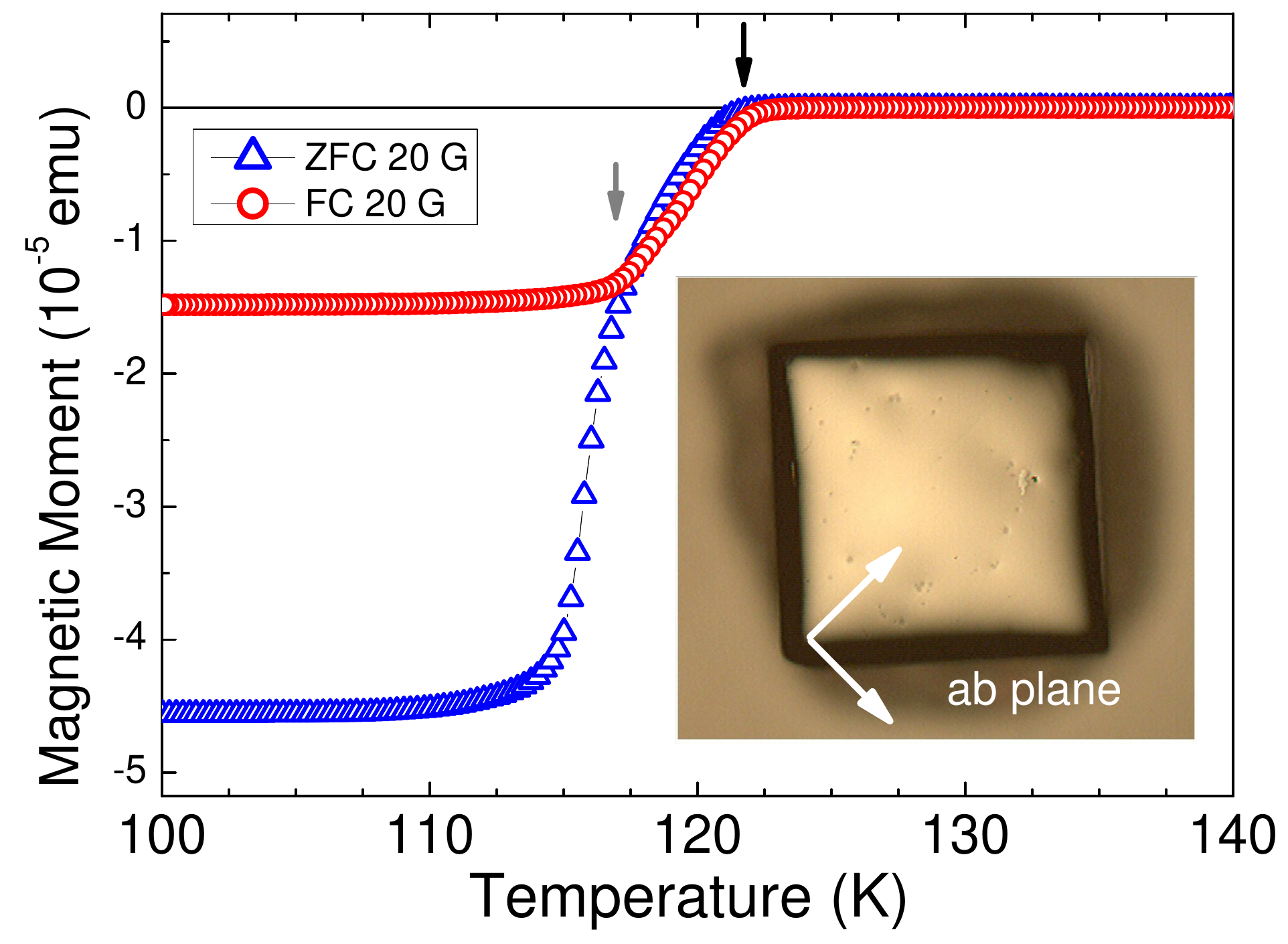}
\caption{Zero field cooling and field cooling magnetizations of under-doped Hg-1223. The applied magnetic field is perpendicular to the (ab) plane and its magnitude is of $\approx20~Gauss$. The black and grey arrows indicate respectively  the $T_{c}$ onset and the middle of the transition.}
\label{fig:squid}
\end{center}\vspace{-5mm}
\end{figure}

The measurements have been performed on slightly underdoped (UD) Hg-$1223$ single crystal grown by a single step synthesis \cite{Colson1994}. The single crystal is a parallelepiped with 0.9 $\times$ 0.9 mm$^2$ cross section and thickness 0.2 mm (see inset of fig.\ref{fig:squid}). The c-axis is normal to the surface with the a-b plane directions 45$\arcdeg$ from the two larger edges. In order to have high optical quality surface, the sample has been polished using diamond paste at $1/10$ \micro m. Dc magnetization measurements under field cooling (FC) and zero field cooling (ZFC) are displayed in fig.\ref{fig:squid}. The $T_{c}$ onset is estimated at 122 K with a middle transition at 117 K. The diamagnetic transition width is $\approx6 K$. The relatively large transition width might reflect slightly doping inhomogeneity \cite{Bertinotti1995}. \\ 

Raman experiments have been carried out using a JY-T64000 spectrometer in single grating configuration using a 600 grooves/mm grating and a Thorlabs NF533-17 notch filter to block the stray light. The spectrometer is equiped with a nitrogen cooled back illuminated 2048x512 CCD
detector. We use the 532 nm excitation line from a diode pump solid state with laser power maintained at 4 mW. Measurements between 10 and 290 K have been performed using an ARS closed-cycle He cryostat. This configuration allows us to cover a wide spectral range ($90~cm^{-1}$ to $4700~cm^{-1}$) with a resolution sets at $5~cm^{-1}$. The complete spectrum for \BAN symmetry is a recombination of three frames centered at different energies. Spectra for other symmetries have been obtained in one frame. Each frame is repeated twice to eliminate cosmic spikes and acquisition time ranged from 5 to 15 minutes. For comparison, in our initial study \cite{Sacuto1997,Sacuto1998,Sacuto2000a} we took almost a day to acquire a spectrum up to $1000~cm^{-1}$. All the spectra have been corrected for the Bose factor and the instrumental spectral response. They are thus proportional to the imaginary part of the Raman response function $\chi^{\prime \prime}(\omega,T)$. Two sets of experimental configurations have been used. In the first set-up the direction of incident electric field is contained in the (ab) plane. The B$_{1g}$ and A$_{1g}$+B$_{2g}$ geometries are obtained from crossed and parallel polarizations, respectively, along the Cu-O bond directions. Then, the crystal is rotated by 45$\arcdeg$ using a Attocube piezo-rotator ANR 101 to obtain the B$_{2g}$ and A$_{1g}$+B$_{1g}$ geometries from crossed and parallel polarizations respectively. The \AO, \BAN and \BN symmetries are related to the $x$ and $y$-components of the Raman susceptibility tensor.  
In the second set-up, the direction of the incident electric field is out of the (ab) plane (along the c-axis). This gives access to the \AO and \E symmetries related to the $z$ and $x$ components of the Raman susceptibility tensor.

\section{EXPERIMENTAL RESULTS}

\begin{figure}[htp!]
\begin{center}
\includegraphics[width=8cm,height=8cm]{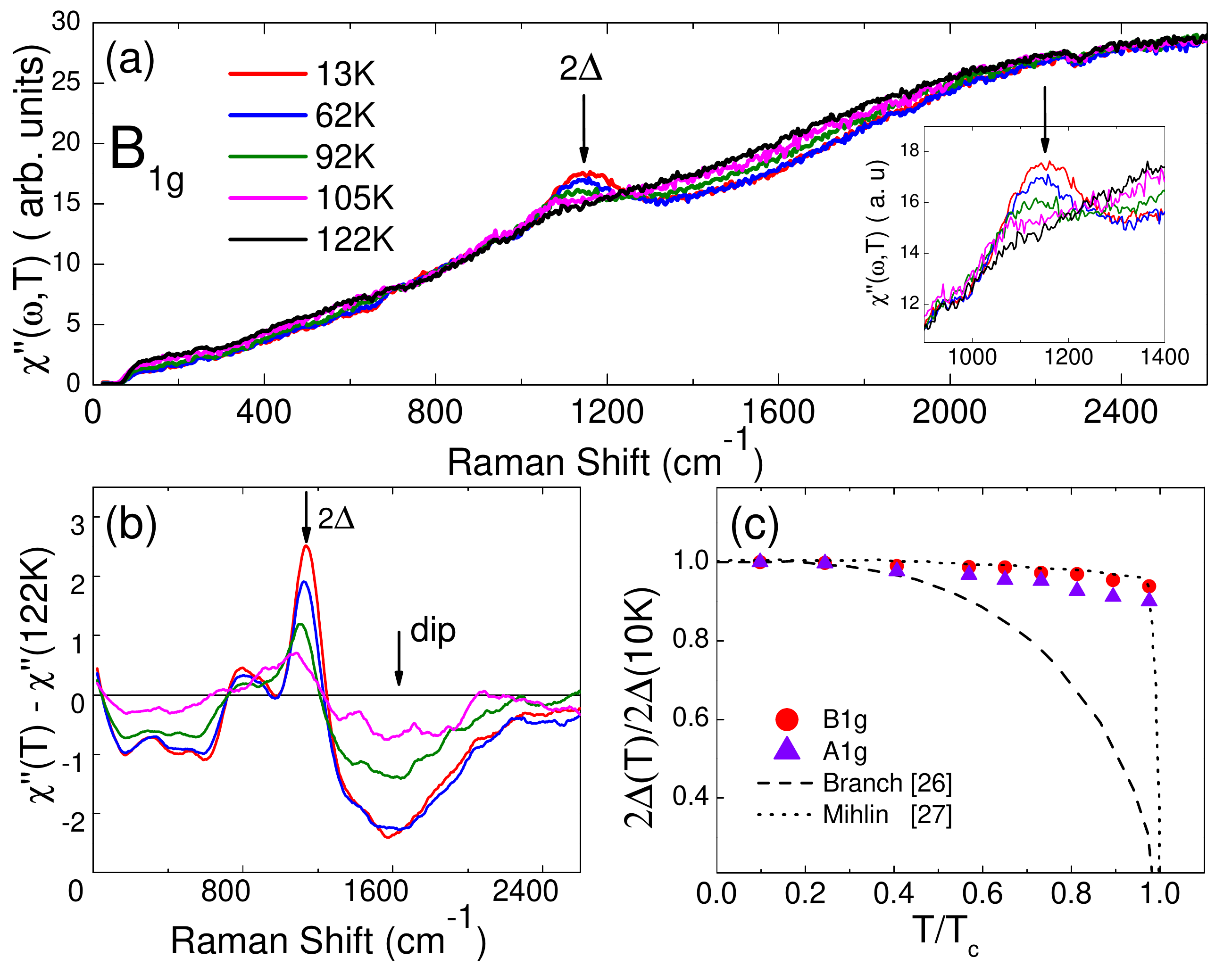}
\caption{(a) Temperature dependence of the \BAN Raman Response up to $T_c$ for the under-doped Hg-1223 single crystal ($T_{c}=122 K$). The $2\Delta$ pairing peak is detected at $\approx1135~cm^{-1}$. (b) Subtraction of the \BAN Raman Responses from the one at $T_{c}=122K$. A peak -dip-hump structure can be observed.  
(c) Temperature dependence of the pairing gap detected in \BAN and \AO symmetries. Its energy remains almost constant up to $T_{c}$ for both symmetries. The dashed and dotted lines correspond respectively to the temperature evolution of a d-wave gap in a weak coupling limit and an order parameter for charge lattice bosons calculation (see text).}
\label{fig:2}
\end{center}\vspace{-7mm}
\end{figure}

In fig.\ref{fig:2}(a) is displayed the temperature dependence of the $B_{1g}$ Raman response $\chi^{\prime \prime}_{\BAN} (\omega, T)$ of the UD Hg-1223 single crystal up to T$_c$= 122 K. We clearly observe the superconducting gap which manifests itself as a relatively sharp quasiparticle pair breaking peak at $2\Delta\approx1135~cm^{-1}$ and with a full width at half maximum of $\approx150~cm^{-1}$ (see inset). The pairing-gap peak growth with decreasing $T$, from the normal to the superconducting state, is associated with the redistribution of the low energy electronic background from below to above $\approx800~cm^{-1}$.  
Interestingly, the pair breaking peak in the SC state is followed by a dip in the electronic continuum around $1600~cm^{-1}$ and a broader hump. The normal (122 K) and superconducting (13 K) electronic background finally merge near $2500~cm^{-1}$.\\

In fig.2 (b) we display the superconducting \BAN Raman responses versus temperature subtracted from the normal one at T$_{c}$, $\chi^{\prime \prime}_{\BAN} (\omega, T)-\chi^{\prime \prime}_{\BAN} (\omega,T_c)$. As the temperature decreases we clearly see the growth of the pairing-gap peak at $1135~cm^{-1}$.  
Notice a weak bump around $800~cm^{-1}$ which was previously erroneously interpreted as the $2\Delta$ pairing peak \cite{Sacuto1997,Sacuto1998,Sacuto2000a}because of the short frequency range ($\approx1000~cm^{-1}$) available. 
Above the pairing gap we detect a dip at $1577~cm^{-1}$ and a hump in the electronic background around $2500~cm^{-1}$. The dip developps immediately below $T_{c}$ as the pairing peak starts to grow. We find $\omega_{dip}/2\Delta\approx1.4$.
This structure has to be compared to the peak-dip-hump reported by ARPES in the spectral function of Bi-2212 at (0,$\pi$) point \cite{Fedorov1999,Campuzano1999,Sato2002,Vishik2010}. Although the electron Raman scattering is a two particles probe and ARPES is a single particle probe, both techniques may be revealing the same electronic structure.  The peak-dip-hump structure observed in Bi-2212 by ARPES is still a matter of debate.  Is it due to an electronic band structure effect or a collective mode ? \cite{Damascelli2003}. The question is still open. It was first interpreted as a bosonic mode due to the coupling between spin fluctuations and electrons \cite{Eschrig2000}  and theoretical Raman investigations suggested that the predominantly linear low frequency dependence of the \BAN Raman response (see fig.2 (a)) and the dip just above the pairing peak hinted at a strong fermionic interaction with spin fluctuations \cite{Chubukov1999}. These features deserve future deeper investigations. 

The temperature dependence of the SC pairing gap is displayed in fig.\ref{fig:2}(c). The critical temperature is sufficiently high to give us an accurate estimate of the the pairing gap energy versus temperature.
It appears that the temperature evolution  of the pairing gap significantly differs from the one expected for a $d-$wave superconducting gap at the antinodes in a weak coupling limit\cite{Branch1995}. The trapezoidal shape is more relevant with a model of charge lattice bosons justified in the strong coupling and short coherence length limit \cite{Mihlin2009,Altman2002,Anderson87}. Such a behavior for the SC gap was also reported by Raman scattering in one and two-copper-oxide-layer compounds slightly underdoped \cite{Guyard2008,Staufer92}.
At low temperature (13 K) we estimate the pairing gap at 14 $k_B$T$_c$ much higher than 4.28 $k_B$T$_c$ expected in the weak coupling limit. This hints at a strong coupling nature for the pairing state. Our findings about the pairing gap energy are in agreement with earlier tunneling and optical measurements on Hg-1223 compounds \cite{Wei1998, McGuire2000}. In tunneling, the gap for an optimally doped Hg-1223 was estimated close to $\sim13$ $k_{B}T_{c}$ and a $d$-wave gap was successfully considered for interpreting the data. In optical conductivity, the pairing gap deduced from the scattering rate was estimated of about 1100 cm$^{-1}$ i.e $\sim13$ $k_{B}T_{c}$.

\begin{figure}[htp!]
\begin{center}
\includegraphics[width=9cm,height=7.5cm]{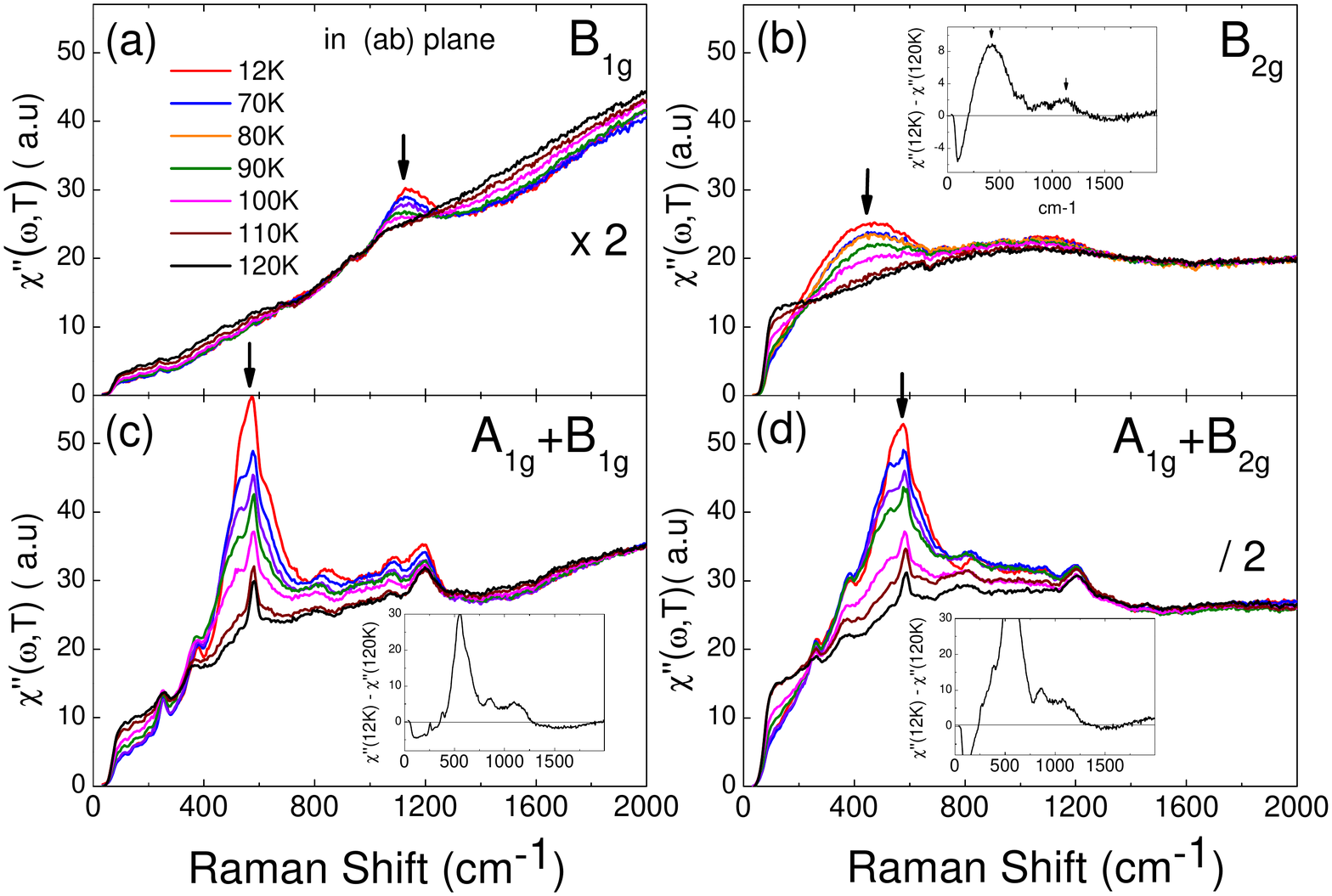}
\caption{Temperature dependencies of the Raman responses in (a) \BAN, (b) \BN, (c) \AO+\BN and (d) \AO+\BAN symmetries obtained from (ab) plane polarizations. The three distinct energy scales can be respectively seen in \BAN, \BN and \AO  at 1135, 430
 and 555 cm$^{-1}$. The insets display the subtracted Raman response between the superconducting (12 K) and the normal (122 K) ones. The insets of panels (c) and (d) have the same intensity scales to facilitate the comparison between the strength of the dip in the electronic continuum.}
\label{fig:3}
\end{center}\vspace{-5mm} 
\end{figure}

In fig.\ref{fig:3} are reported the Raman spectra obtained from light polarizations in the (ab) plane. The temperature dependencies of the Raman responses $\chi^{\prime \prime}_{\nu} (\omega, T)$ for $\nu = B_{1g}, B_{2g}, A_{1g}+B_{1g}, A_{1g}+B_{2g}$ symmetries are displayed up to $T_c$. The Raman spectra in B$_{1g}$ and B$_{2g}$ symmetries (panels (a) and (b)) are free of phonon structures. This allows us to study the electronic background in a reliable way. In the A$_{1g}$+B$_{1g}$ and  A$_{1g}$+B$_{2g}$ spectra (panel (c) and (d)), we observe sharp intensive phonon lines at 256, 380 and 591 cm$^{-1}$. A broader peak is detected at higher energy close to 1180 cm$^{-1}$ which might reflect a double phonon structure stemming from the 591 cm$^{-1}$ phonon line.


Focusing on the electronic background we detect three distinct electronic peaks in the SC state of Hg-1223 depending on the symmetry. The peak at 1135 cm$^{-1}$  detected in the \BAN spectrum (panel (a)) has already been assigned to the $d$-wave pairing gap (see above). 
In \BN spectrum (panel (b)) we observe a broad electronic peak at 430 cm$^{-1}$, whose full width at half maximum (FWHM) is $\approx 2.5$ larger than the \BAN peak (see inset), and a small bump around 1135 cm$^{-1}$. We can see that the intensity of the \BN superconducting peak is stronger than the \BAN one. This has already been observed for underdoped mercurate compounds \cite{Letacon2006}. The d-wave gap amplitude is maximum at the antinodes (\BAN symmetry) while it vanishes at the nodes (\BN symmetry). Therefore, as expected, the energy of the \BAN peak is larger than the \BN one. 
However the energy ratio between the \BAN and \BN peak energies is particularly high $\approx2.6$. This is hard to reconcile with a d-wave SC gap which develops on a full cylindrical Fermi surface as in this picture the ratio would be close to 1.3 \cite{Devereaux1995}. A possible explanation is to consider the formation of Fermi arcs \cite{Norman1998}, with a quasiparticles spectral weight mostly concentrated around the nodal regions. This can push the \BN peaks to lower energy, as discussed in previous works \cite{Chen1997,Chubukov2008,Leblanc2010,Blanc2010,Sacuto2013} and explain such a high  \BAN/\BN energy ratio. 

In the \AO+\BAN spectrum (panel (c)) we also observe an intense SC electronic peak at 555 cm$^{-1}$ (called the \AO peak) whose FWHM is sharp enough (unlike the \BN one) to track its temperature evolution (see fig. 2 (a)). It shifts slowly toward low energy as the temperatures raises in a similar way to the \BAN pairing peak. This hints that the \AO peak is connected to the $2\Delta$ pairing peak. In other hole-doped cuprates such as Bi-2212, the \AO peak exhibits the same doping dependence than the \BAN pairing peak \cite{Benhabib2015b}. These results suggest that the \AO peak is likely connected to a collective mode located below the $2\Delta$- threshold of the particle-hole continuum \cite{Venturini2000,Gallais2002,Bourges2005,Montiel2015}. The \AO peak detected in the \AO+\BN spectrum (fig.3(d)) has a FWHM much larger than the one in the \AO+\BAN spectrum (fig.3(c)). This is due to the additional broadening stemming from the \BN peak which is $\approx 2.5$ larger than the one of the \BAN peak \cite{Note}.

We have found in fig.\ref{fig:2} (b) that the electronic background of the \BAN spectrum exhibits in the SC state a dip centered $\approx1577$ cm$^{-1}$ just above the pairing gap. We can now estimate (from the subtracted Raman responses in fig.2 (b) and in the inset of fig.3 (b))
that the dip magnitude is higher by a factor of 4 in the \BAN spectrum in comparison with the \BN one. The same effect in the dip magnitude is also observed between the \AO+\BAN and \AO+\BN spectra. See for comparison insets of fig.3 (c) and (d). This confirms that the dip is a characteristic feature of the \BAN Raman spectra and mainly relevant to antinodal electronic properties as also noticed in Bi-2212 \cite{Damascelli2003}.

\begin{figure}[htp!]
\begin{center}
\includegraphics[width=8cm,height=7.5cm]{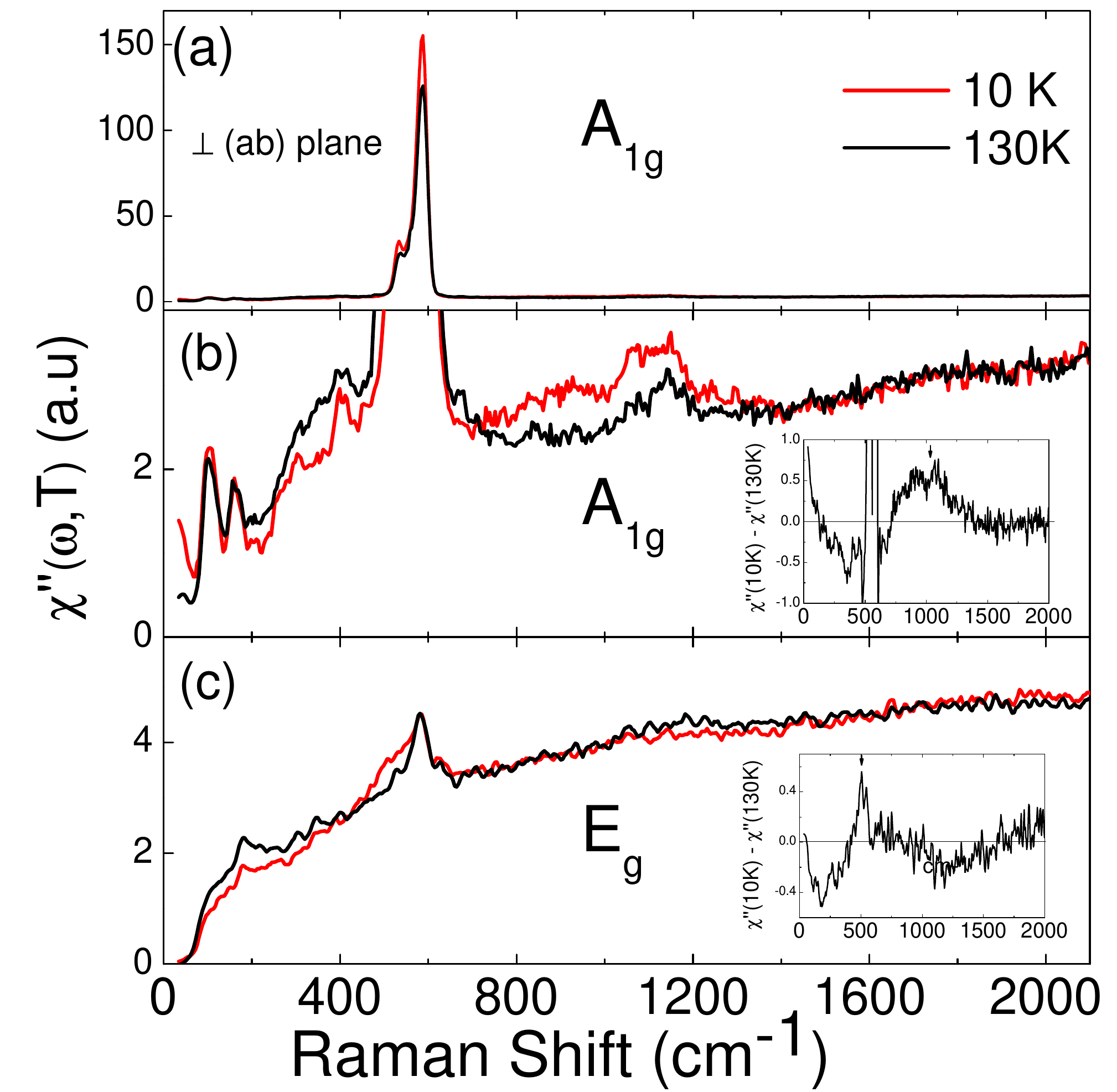}
\caption{Superconducting and normal Raman responses in (a, b): \AO and (c): \E symmetries obtained from c-axis polarizations}
\label{fig:4}
\end{center}\vspace{-5mm}
\end{figure}

We finally turn to the Raman measurements with light polarization perpendicular to the (ab) plane i.e: along the c-axis. In fig.\ref{fig:4} are displayed the Raman spectra in \AO and \E symmetries which correspond respectively to parallel and crossed polarizations (cf. experimental procedures). In \AO symmetry (panels (a) and (b)), we detect a huge phonon line at 585~cm$^{-1}$ with a shoulder at 534 cm$^{-1}$. These peaks are respectively assigned to the vertical motions of the apex oxygen atoms in the BaO plane and interstitial oxygen atoms in the HgO plane \cite{Sacuto1996,Zhou1996, Zhou1997,Zhou1997bis}. 
The very strong Raman activity of the 585 cm$^{-1}$ mode can be explained by (i) the polarizability enhancement due to the same direction of the atomic motions and the electric field along the c-axis and (ii) the phonon lifetime increasing along the c-axis due to the loss quasiparticle spectral weight perpendicular to the (ab) plane. Indeed the temperature dependence of the c-axis resistivity exhibits a semiconductor like behavior unlike the (ab) plane resistivity (which exhibits a metallic one) and a strong anisotropy $\rho_{c}\approx1000~\rho_{ab}$ \cite{Carrington1994}. Although the electronic continuum is weaker along the c-axis than in the (ab) plane, we detect in \AO symmetry (panel (b)) a redistribution of the electronic background from low to high energy and the growth of the pairing gap as the temperature decreases below $T_{c}$ . After subtracting the \AO normal Raman response at 130 K from the superconducting one at 10 K, (see inset of panel (b)), we find a SC peak at 1135 cm$^{-1}$, i.e. the same energy of the pair breaking peak detected in \BAN symmetry (see fig.3 (a)). 
Probing the quasiparticles dynamics along the c-axis is thus equivalent to probe the \BAN quasi-particles dynamics in the (ab) plane. This can be understood by noticing that the square of the \BAN Raman vertex $(cosk_x-cosk_y)^{2}$ has the same symmetry as the c-axis hopping \cite{Andersen1995}. Our results confirm the close relationship between the antinodal electronic properties in the (ab) plane and the c-axis ones.  
In \E symmetry (panel (c))  the subtraction of the normal to the superconducting Raman responses reveals a weak SC peak around 555 cm$^{-1}$ close in energy to the one already detected in \AO symmetry (see fig.3 (c)).  

In conclusion, by performing electronic Raman scattering on Hg-1223 single crystal over a wide energy range we reveal spectral features previously overlooked. We succeeded to detect three distinct SC electronic peaks at 1135, 555  and 430 cm$^{-1}$ respectively in the \BAN, \AO and \BN symmetries, obtained from in and out of plane polarizations. The 1135 cm$^{-1}$ peak is assigned to the d-wave pairing gap at the antinodes. Its huge energy value $\approx$ 14 $k_{B}T_{c}$  
and its weak temperature dependence below $T_{c}$ hints at a strong coupling pairing state. 
Remarkably the 1135 cm$^{-1}$ peak is also detected in \AO symmetry along the c-axis which confirms the intimate connection between the quasiparticles dynamics at the antinodes within the (ab) plane and along the c-axis in the SC state.
Finally, we find that the 1135 cm$^{-1}$ superconducting peak displays a peak-dip-hump structure in the electronic background of the antinodal Raman spectrum similar to the one detected by ARPES in Bi-2212. Identify its origin requires a more deeper study which will be addressed in a near future. 

We are grateful to M. Civelli, I. Paul, A. Auerbach and A. Georges for fruitful discussions. Correspondences and requests for materials should be addressed to A.S. (alain.sacuto@univ-paris-diderot.fr) and B.L. (bastien.loret@univ-paris-diderot.fr)

\bibliography{rapidcom}

\begin{thebibliography}{45}
\bibitem [{Not()}]{Note}%
  \BibitemOpen
  \href@noop {} {\bibinfo  {journal} {Notice that the electronic background
  level of the A$_{1g}$+ B$_{2g}$ spectrum below 1000 cm$^{-1}$ is higher than
  the A$_{1g}$+ B$_{1g} $ one. This is a consequence of a stronger Raman
  intensity of the B$_{2g} $ peak compared to the B$_{1g} $ one.}\
  }\BibitemShut {NoStop}%
\end{thebibliography}

\end{document}